\documentclass{mem}
\usepackage[comma,authoryear]{natbib}\usepackage{txfonts}\usepackage{balance}
\usepackage{graphicx}
\usepackage[a4paper,breaklinks,dvipdfm]{hyperref}
\idline{75}{282}
\begin{document}
\def\teff{$T\rm_{eff }$}
\def\kms{$\mathrm {km s}^{-1}$}

\def\ni{\noindent}
\def\lsim{\mathrel{\rlap{\lower4pt\hbox{\hskip1pt$\sim$}}
    \raise1pt\hbox{$<$}}}
\def\gsim{\mathrel{\rlap{\lower4pt\hbox{\hskip1pt$\sim$}}
    \raise1pt\hbox{$>$}}}
\def\sqr#1#2{{\vcenter{\vbox{\hrule height.#2pt
         \hbox{\vrule width.#2pt height#1pt \kern#1pt
         \vrule width.#2pt}
         \hrule height.#2pt}}}}
\def\square{\mathchoice\sqr66\sqr66\sqr{2.1}3\sqr{1.5}3}

\def\beq{\begin{equation}}
\def\eeq{\end{equation}}
\def\beqa{\begin{eqnarray}}
\def\eeqa{\end{eqnarray}}

\def\to{\rightarrow}
\def\gev{\mbox{GeV}}
\def\ev{\mbox{eV}}
\def\mev{\mbox{MeV}}
\def\tev{\mbox{TeV}}
\def\m{\mbox{cm}}
\def\mpc{\mbox{Mpc}}

\title{
What if ... General Relativity is not the theory?
}

   \subtitle{}

\author{O. \,Bertolami\inst{1,2} }

  \offprints{O. Bertolami}

\institute{
Departamento de F\'\i sica e Astronomia  \\
Faculdade de Ci\^encias, Universidade do Porto \\ 
Rua do Campo Alegre 687, 4169-007 Porto, Portugal.
\and
Instituto de Plasmas e Fus\~ao Nuclear,
 Instituto Superior T\'ecnico, 
Avenida Rovisco Pais 1, 1049-001 Lisboa, Portugal.
\email{orfeu.bertolami@fc.up.pt}
}

\authorrunning{Bertolami}

\titlerunning{What if ... General Relativity is not the theory?}

\abstract{
The nature of gravity is fundamental to understand the scaffolding of the Universe and its evolution. Einstein's general theory of relativity has been scrutinized for over ninety five years and shown to describe accurately all phenomena from the solar system to the Universe. However, this success is achieved in the case of the largest scales provided one admits contributions to energy-momentum tensor involving dark components such as dark energy and dark matter. 
Moreover, the theory has well known shortcomings, such as the problem of singularities, the cosmological constant problem and the well known initial conditions problems for the cosmological description. Furthermore, general relativity also does not fit the well known procedures that allow for the quantization of the other fundamental interactions. In this discussion we briefly review the experimental bounds on the foundational principles of general relativity, and present three recent proposals to extend general relativity or, at least, to regard it under different perspectives.
}
\maketitle{}

\section{Introduction}

The understanding of  the Universe in its challenging complexity requires a detailed knowledge of the force of gravity at all scales. 
Gravity was the first of the known four fundamental interactions of Nature to be described through a detailed theoretical framework.  
Newton was the first to understand that gravity was universal, dictating the fall of any body on Earth, as well as the planetary motion in the solar
system.  On larger scales the role of gravity is even
more overwhelming and it drives the evolution of the galaxies, clusters and superclusters of galaxies and ultimately the whole dynamics 
of the Universe. Currently, Einstein's general relativity (GR) allows for a detailed description of a vast range of phenomena, from the
dynamics of compact astrophysical objects such as neutron stars to cosmology, where the Universe itself is the
object of study. The list of predictions of GR is quite impressive and includes, for instance, the expansion of the Universe, gravitational
lensing, gravitational waves, and black holes. Remarkably, only gravitational waves and black holes have not been directly and unquestionably detected. 
And yet, despite of all that, little is known about the underlying microscopic theory that supports the fundamental connection between space-time curvature and 
the energy-momentum of matter as prescribed by GR.  

In what concerns the Universe, GR is up to the job as far as the contribution of ``dark" components, dark energy and dark matter, to the energy-momentum tensor is included. Furthermore, GR seems to be incomplete as it is unable to address issues such as the problem of singularities, the cosmological constant problem and the need of extra states to play the role of the inflaton field so to avoid well known initial conditions problems for the cosmological description. 

In what follows we briefly review the experimental bounds on the foundational principles of  GR, and discuss three 
recent proposals to extend the theory and to regard it under alternative perspectives. These three approaches have been chosen by their ``freshness"  and because they emerge from new underlying assumptions about scale invariance, the thermodynamic features of gravity and the holographic principle, and Lorentz symmetry.   

This contribution is organized as follows: in the next section, we briefly review the main bounds on the foundations of GR, as well as on the CPT symmetry; in section III, we discuss the unparticle proposal and its ``ungravity" connection; in section IV, we discuss a putative relationship among Newton's law, thermodynamics, the holographic and the equivalence principles; in section V, we discuss the Ho\v{r}ava-Lischitz proposal for quantum gravity and speculate on some possible implications for its  quantum cosmology. Some final remarks are advanced at the last section.

\section{The experimental state of the foundational principles of GR}

We briefly review now the current experimental bounds on the underlying principles of GR. 
They are the following (see \citep{Will2005,BPT2006} for thorough discussions):

\vskip 0.3cm
\ni
1. {\it Weak Equivalence Principle} (WEP), which states that 
freely falling bodies do have the same acceleration 
in the same gravitational field, independent on their compositions. In the laboratory, precise tests of the WEP are performed by comparing
the free fall accelerations, $a_1$ and $a_2$, of different test
bodies at the same distance from the source
of the gravity as a measure of the difference between the ratio of the gravitational mass, $M_G$, to the inertial mass, $M_I$, of each body:
\beq {\Delta a \over a} = {2(a_1- a_2) \over a_1 + a_2} =\left({M_G \over M_I}\right)_{1} -\left({M_G \over M_I}\right)_{2}~.
\label{WEP_da} 
\eeq
The most recent tests yield $\Delta a/a \lsim10^{-13}$ \citep{Adelberger}.

\vskip 0.3cm
\ni
2. {\it Local Lorentz Invariance} (LLI), implies that 
clock rates are independent on the 
clock's velocities. Thus, invariance under Lorentz transformations implies that the laws of
physics are independent of the velocity of the reference frame . Limits on the violation of Lorentz symmetry are available from
laser interferometric versions of the Michelson-Morley experiment,
by comparing the speed of light, $c$ and the maximum attainable
velocity of particles, $c_i$, up to $\delta \equiv |c^2/c_{i}^2 - 1| < 10^{-9}$ \citep{Brillet}. Limits on Lorentz
symmetry can also be obtained from the absence of high-energy cosmic
rays with energies beyond $5 \times 10^{19}$~eV from sources 
further than 50--100~$\mpc$  \citep{Abbasi,Abraham et al}, the so-called
Greisen-Zatsepin-Kuzmin (GKZ) cut-off \citep{Greisen,Zatsepin-Kuzmin}. As discussed in Refs. 
\citep{Sato,Coleman,Mestres,Bertolami2} slight
violations of Lorentz invariance would cause energy-dependent
effects that would suppress otherwise inevitable processes such as
the hadronic resonant production of, for instance, $\Delta_{1232}$, due to the scattering of ultra-high energy protons with photons of the 
cosmic microwave background radiation. The study of the kinematics of this process yields
a quite stringent constraint on the Lorentz invariance, namely 
$\delta \simeq 1.7 \times 10^{-25}$ \citep{Bertolami2,Bertolami6}.

\vskip 0.3cm
\ni
3. {\it Local position invariance} (LPI), postulates that clocks rates are also independent 
on their space-time positions. 
Given that both, WEP and LLI, have been tested with
great accuracy, experiments concerning the universality of the
gravitational red-shift of frequencies, $\nu$, measure the extend to which the LPI holds.
Therefore, violations of the LPI would imply that the rate of a
free falling clock would be different when compared with a
standard one, for instance on Earth's surface. The accuracy to
which the LPI holds can be
parameterized as $ \Delta \nu / \nu = (1 + \mu) U / c^2$, where $U$ is the Newtonian gravitational potential. 
An accurate verification of the LPI was achieved by comparing hydrogen-maser frequencies on Earth
and on a rocket flying to an altitude of $10^4$~km  \citep{Vessot}, yields $ \vert \mu \vert < 2 \times 10^{-4}$, 
an experiment usually referred to as Gravity Probe-A.
Recently, an order of magnitude improvement was attained, $ \vert \mu \vert < 2.1 \times 10^{-5}$ \citep{Bauch}.

\vskip 0.3cm
\ni
The {\it Strong Equivalence Principle} (SEP) extends the WEP to the gravitational energy itself. GR assumes that the SEP is exact,
but alternative metric theories of gravity such as those involving
scalar fields and other extensions of gravity theory, violate the SEP. 

Putative violations of the SEP can be expressed in the PPN formalism (see \citep{Will1993} and refs. therein) by the ratio between
gravitational and inertial masses, $M_G/M_I$ \citep{Nordtvedt_1968a}: 
\beq \left[{M_G \over M_I}\right]_{\tt SEP} = 1 +\eta\left({\Omega \over Mc^2}\right), \label{eq:MgMi} \eeq
\noindent 
where $M$ is the mass of a body, $\Omega $ is the body's
(negative) gravitational self-energy (see below), $Mc^2$ is its total
mass-energy, and $\eta$ is a dimensionless constant that parametrizes SEP
violations \citep{Nordtvedt_1968a}. In a 
fully-conservative, Lorentz-invariant theories of gravity
\citep{Will1993} the SEP parameter is related with the PPN
parameters by $ \eta = 4\beta - \gamma -3$. In GR
$\beta = 1$ and $\gamma = 1$, so that $\eta = 0$, where the PPN parameters are introduced through the expansion of the 
metric as {}
\beqa g_{00}&=&-1+2U - 2\beta\, U^2 + ... ~, \\ \nonumber g_{0i} & = &-{1\over 2}(4\gamma+3)V_i + ... ~, \\ \nonumber
g_{ij}&=&\delta_{ij}(1+2\gamma U)+ ...~~,
\label{eqno(1)} \eeqa
where $V_i$ is the component of the velocity of a test particle and $U$ the Newtonian gravitational potential. 

The self energy of a body $B$ is given by
\beq \left({\Omega  \over Mc^2}\right)_B = - {G \over 2 M_B
c^2}\int_B d^3{\bf x} d^3{\bf y} {\rho_B({\bf x})\rho_B({\bf y})
\over | {\bf x} - {\bf y}|}. \label{eq:omega} \eeq
\noindent 
For a sphere with a radius $R$ and uniform density,
$\Omega /Mc^2 = -3GM/5Rc^2$. Accurate evaluation for solar system bodies
requires numerical integration of the expression of
Eq.~(\ref{eq:omega}). Given that the gravitational self-energy is proportional to
$M^2$ and the extreme weakness of gravity, the
typical values for the ratio $(\Omega /Mc^2)$ are $O(10^{-25})$
for laboratory size bodies. Therefore, to test the SEP one must consider planetary-sized
bodies, where the ratio Eq.~(\ref{eq:omega}) is
considerably higher. To date, the Earth-Moon-Sun system provides the most accurate
test of the SEP: recent results from lunar laser ranging (LLR) data yield $\Delta (M_G/M_I)_{\tt WEP}
=(-1.0\pm1.4)\times10^{-13}$ \citep{Williams_Turyshev_Boggs_2004}.
This result corresponds to a test of the SEP of $\Delta
(M_G/M_I)_{\tt SEP} =(-2.0\pm2.0)\times10^{-13}$ with the SEP
violation parameter found to be: 
$\eta=(4.4\pm4.5)\times 10^{-4}$. Using the recent Cassini result
for the PPN parameter $\gamma-1=(2.1\pm2.3)\times  10^{-5}$ \citep{cassini_ber}, PPN parameter $\beta$ is
determined at the level of $\beta-1=(1.2\pm1.1)\times 10^{-4}$.

In quantum field theory, Lorentz invariance together with the requirements of locality, an Hermitian Hamiltonian and the spin-statistics connection give rise to the CPT theorem (see e.g Ref. \citep{Weinberg}) which implies that the mass of particles and their anti-particles are exactly the same. This equality can be measured with great accuracy for the $K^0-\overline{K^0}$ system, and it yields the bound $8 \times 10^{-19}$ in terms of their averaged mass of the system \citep{PDG2011}. 

The discussion of the fundamental underlying symmetries and principles of GR  leads us to a broader discussion of the various alternative theories, and hence, also to raise questions about some crucial theoretical problems and experiments. A partial list should include, for instance: 

\vskip 0.3cm
\ni
Can dark matter (DM) be direct detected and its nature unravelled?

\vskip 0.3cm
\ni
Can one determine the nature of dark energy (DE) solely from the determination of the equation of state of the Universe? Does DE interact with DM? Can DE be described in an unified way together with DM 
\citep{Kamenshchik2001,Bilic2002,Bento2002,Barreiro2008}?

\vskip 0.3cm
\ni
Do Pioneer \citep{Anderson2002} and flyby \citep{Anderson2008} anomalies really hint about new physics? Can thermal effects account for the Pioneer anomaly as recent calculations suggest \citep{BFGP2008}? Can one test the flyby anomaly \citep{BFGP2011}?

\vskip 0.3cm
\ni
Can one expect violations of the WEP at cosmological scales \citep{BGL2007}? Is there any hint of a connection between a violation of the WEP and the cosmological 
constant problem \citep{OB2009}?

\vskip 0.3cm
 \ni
How realistic is to expect that the detection of gravitational waves and black holes will allow for testing GR against alternative theories of gravity?

\vskip 0.3cm
\ni
Are there new forces of nature? New symmetries to be accounted for? Can one connect putative variations of the fundamental constants of nature with the existence of extra dimensions?

\vskip 0.3cm
\ni  
Does space-time has a noncommutative structure \citep{Connes,Seiberg}? Should noncommutativity be extended to the phase space (see e.g. Ref. \citep{Bastos})?

\vskip 0.3cm
\ni
Is gravity an emergent or a fundamental interaction?

\vskip 0.3cm
\ni 
Of course, we could connect this discussion to the quest of unifying the interactions of nature and to the task of quantizing gravity, but this would lead us to the vast and well studied grounds of string theory, braneworlds and so on. We will not pursue this path here, but instead, discuss three approaches where some of the topics raised above are addressed in a quite concrete way, namely: ungravity, emergent gravity and its connection with phase space noncommutativity, and the Ho\v{r}ava-Lifshitz gravity proposal and its quantum cosmology.  

\section{{\bf Ungravity}}

A quite interesting issue concerning gravity and the Standard Model (SM) of the electroweak and strong interactions is the existence of yet undetected hidden symmetries. It has been 
conjectured that the SM could signal an underlying scale invariant, through the presence of states dubbed unparticles \citep{georgi}. 
Implementing scale invariance requires considering an additional set of fields with a nontrivial infrared (IR) fixed point, the so-called Banks-
Zacks (BZ) fields. The interaction between SM and BZ fields would occur through the exchange of particles with a 
large mass scale, $M_*$, which can be written as
\beq {\cal L}_{BZ} = {1\over M_*^k}O_{SM}O_{BZ}~~,
\label{BZ}
\eeq \noindent where $O_{SM}$ is an operator with mass dimension $d_{SM}$ built out of SM fields and 
$O_{BZ}$ is an operator with mass dimension $d_{BZ}$ built out of BZ fields.

At an energy scale $\Lambda_U$ the BZ operators match onto unparticles operators ($O_{U}$) 
and Eq. (\ref{BZ}) matches onto
\beq \label{LagU}
{\cal L}_U={C_U \Lambda_U^{d_{BZ}-d_U}\over M_*^k}O_{SM}O_{U}~~,
\eeq \noindent where $d_U$ is the scaling dimension of the unparticle operator $O_U$, which can be fractional, and $C_U$ is a coefficient function.

These new states may have a bearing on gravity whether there exists tensor-type unparticle interactions, which would couple to the stress-energy tensor of SM states and modify the 
Newtonian potential $U(r)$. This modification, usually referred to as ``ungravity", leads to a power-law addition to the Newtonian potential \citep{ungravity}
\beq V(r)=-{G_U M \over r}\left[1+\left({R_G \over r}\right)^{2d_u-2}\right]~~,
\label{UPO}
\eeq \noindent where $R_G$ is the characteristic length scale of ungravity,
\beqa 
\label{R_G} R_G & =& {1\over \pi \Lambda_U} \left({M_{Pl} \over M_*}\right)^{1/(d_U-1)} \times \\ \nonumber && \left[ {2(2-
\alpha) \over \pi} {\Gamma(d_U+{1 \over 2})\Gamma(d_U-{1 \over 2}) \over \Gamma(2d_U)}\right]^{1/
(2d_U-2)}~~,
\eeqa 
\noindent and $\Lambda_U > 1~TeV$ is the energy scale of the unparticle interaction (the lower bound 
reflects the lack of detection of these interactions within the available energy range), $M_{Pl} $ is the Planck 
mass and $\alpha$ is a constant dependent on the type of propagator (unity in the case of a graviton).

The Newtonian potential is recovered for $d_U = 1$, $R_G = 0$ (if $d_U>1$) or 
$R_G \rightarrow \infty$ (if $d_U<1$), so that
\beq G_U = {G \over 1+\left({R_G \over R_0}\right)^{2d_U-2}}~~,
\label{G-G_N}
\eeq \noindent where $R_0$ is the distance where the 
gravitational potential matches the Newtonian one, $V(R_0 ) = U(R_0)$. Given that the value of $R_0$ 
is unknown, only values of $d_U$ close to the unity are considered \citep{ungravity}, so that 
Eq. (\ref{UPO}) is approximately given by
\beq V(r)=-{G M \over 2r}\left[1+\left({R_G \over r}\right)^{2d_u-2}\right]~~.
\label{UP}
\eeq 
\noindent Notice that corrections of this type also arise in the context of a gravity model with vector-
induced spontaneous Lorentz symmetry breaking \citep{bumblebee}.

There has been several attempts to constrain ungravity models and ungravity inspired deviations from Newton's law. These include studies of the 
hydrostatic equilibrium conditions of a star and the ensued perturbation of the Lane-Emden equation, a method developed in Ref. \citep{method} and used, 
in Ref. \citep{range}, to constrain $f(R)$ gravity theories 
with non-minimal coupling between curvature and matter \citep{BBHL2007}. 
From the resulting change in the star's central temperature, we obtain constraints on the ungravity parameters $R_G$ and $d_U$. 
Given that the overall properties of the 
Sun are well described by the $n=3$ polytropic index and that the allowed uncertainty on Sun's central temperature is $\Delta T_c / T_c \approx 0.06$, 
we find for $d_U \gtrsim 1$, $R_G \lsim R_{Sun} \simeq 7 \times 10^8$ m and $\Lambda_U \geq 1~TeV$, lower bounds on $M_*$ are in the range 
$(10^{-2}-10^{-1})M_{Pl}$. For $d_U \lesssim 1$, $R_G \gtrsim R_{Sun}$ and $\Lambda_U \geq 1~TeV$, $M_*$ must lie in the range 
above $(10^{-1}- 10^2)M_{Pl}$ \citep{BPS2009}. 

These results for $d_U \gtrsim 1$ are either more stringent or similar to those previously available \citep{cosmo0,freitas,hsu,das,cosmo}. 
The lower bound derived for $d_U \lesssim 1$ is more relevant, since this was unexplored. Complementary bounds can be 
obtained by examining the effect of ungravity on nucleosynthesis yields \citep{OBNS}.

\section{{\bf Emergent Gravity}}

Recently, it has been argued that Newton's inverse square law (ISL) and GR can be derived from thermodynamical considerations and from the holographic principle \citep{Verlinde}. In this formulation, the ISL arises from 
somewhat more fundamental
entities such as energy, entropy, temperature and the counting of degrees of freedom set by the holographic principle \citep{tHooft,Susskind}. Thus, the crucial  issue in the emergence of gravity, as in the case of black holes, is the relationship between information and entropy. The variation of entropy is assumed to be generated by the displacement of matter, leading to an entropic force, which eventually takes the form of gravity. 

An implication of this entropic derivation of the ISL is that gravity is not a fundamental interaction, or at least, it is not directly related with the spin-2 state associated with the graviton, encountered in fundamental theories such as string theory. Actually, these arguments are in line with earlier ideas about the emergence of gravity from thermodynamical considerations \citep{Jacobson,Padmanabhan}. Central in the arguing of  Ref. \citep{Verlinde} are the holographic principle \citep{Maldacena,Bousso} and the Bekenstein bound between the variation of the entropy, $\Delta S$ and the relativistic energy, $E$ \citep{Bekenstein}: 
$\Delta S \le 2 \pi k_B E \Delta x / \hbar c$, in terms of Planck's and Boltzmann's constants, the speed of light and a matter displacement, $\Delta x$. These arguments can be extended to retrieve GR \citep{Verlinde}. 

In Ref. \citep{Bastos3a} we have argued that given the relevance of phase-space noncommutativity for quantum cosmology \citep{Bastos1}, for the thermodynamics of the Schwarzschild black hole (BH) \citep{Bastos2} and for the BH singularity problem \citep{Bastos3,Bastos3b}, it is just natural that this more general form of noncommutativity is considered when analyzing the emergence of gravity.

The starting point is the phase space noncommutative algebra for configuration and momentum variables in $d$ space dimensions:

\beq\label{eq1.1}
\left[\hat q'_i, \hat q'_j \right] = i\theta_{ij}, \hspace{0.01 cm} \left[\hat q'_i, \hat p'_j \right] = i \hbar \delta_{ij},
\hspace{0.01 cm} \left[\hat p'_i, \hat p'_j \right] = i \eta_{ij}, 
\eeq
where $ i,j= 1, ... ,d$, $\eta_{ij}$ and $\theta_{ij}$ are antisymmetric real constant ($d \times d$) matrices and $\delta_{ij}$ is the identity matrix. 

Repeating the entropic derivation of the ISL in the context of phase space noncommutativity leads, for $\theta\eta / \hbar^2 << 1$, to Newton's law and a noncommutative correction \citep{Bastos3a}:
\beq\label{Eq2.17}
F_{NC}= {GMm\over r^2} \left(1+ {\theta_{12}\eta_{12}\over2 \hbar^2} \right)~.
\eeq

This noncommutative correction is not isotropic, which affects the gravitational fall of masses on different directions. The expected  relative differential acceleration is given, if one considers, say fall along the plane $(1,2)$ with acceleration, $a_1$, and say along the plane $(2,3)$ with acceleration $a_2$:
\beq\label{Eq3.1}
{\Delta a\over a} = 2\left({a_1-a_2 \over a_1+a_2}\right) \simeq {1\over 4\hbar^2}(\theta_{12}\eta_{12} - \theta_{23}\eta_{23}).
\eeq
Therefore, as it stands, this result sets a bound on the anisotropy of noncommutativity. If however, we further assume that $\theta_{12}\eta_{12} \simeq O(1) \theta_{23}\eta_{23} \equiv \theta\eta$, then the above mentioned bound on the WEP implies for the dimensionless quantity $\theta\eta / \hbar^2$:
\beq
{\theta\eta\over \hbar^2} \lsim O(1) \times 10^{-13}~.
\label{Eq3.2}
\eeq
This bound states that noncommutative effects can be 10 orders of magnitude greater than discussed in the context of the noncommutative gravitational quantum well \citep{BRACZ05} and the Lamb shift correction to the relativistic hydrogen atom \citep{BQ2011}. However, this result does not presuppose any value for the $\theta$ parameter. This bound implies that if, for instance, $\sqrt{\theta} \lsim (10~TeV)^{-1}$ as inferred from the induced Lorentz invariance in the electromagnetic sector of the SM extension due to configuration space noncommutativity \citep{Carroll}, then $\sqrt{\eta} \gsim 10^{-4}~GeV$. The latter bound is unsatisfactory as it implies that noncommutative effects should already have been observed. If however, one assumes that $\sqrt{\theta} \gsim M_P^{-1}$, i.e. that the characteristic scale of $\theta$ is essentially the scale of quantum gravity effects (the Planck mass, $M_P=L_P ^{-1}$), then $\sqrt{\eta} \gsim 10^{-6}~M_P$, an interesting intermediary scale. 
 At this point, further advancement is only possible with new  experimental data, either by the direct identification of noncommutative effects or through more stringent bounds on the WEP.

A surprising implication of the above result is the connection with the observed value of the cosmological constant. Indeed, it has been argued that a putative breaking of the WEP at about $\Delta_{WEP} \simeq10^{-14}$ has a bearing on the discrepancy between the observed value of the cosmological constant and the expected one from the SM electroweak symmetry breaking \citep{OB2009}:
\beq
{\Lambda_{Obs.} \over \Lambda_{SM}} = {\Delta^4_{EP}}~,
\label{Eq4.1} 
\eeq
and thus
\beq
{\Lambda_{Obs.}\over \Lambda_{SM}} \simeq \left({\theta\eta\over \hbar^2}\right)^4 ~.
\label{Eq4.2}
\eeq

Thus, the emergence of gravity through thermodynamical arguments, besides its own intrinsic pertinence, provides a suggestive connection between the parameters of phase-space noncommutativity with the observed value of the cosmological constant.
Of course, this relationship holds as far as the equivalence principle is found to be violated at $\Delta_{EP} \simeq10^{-14}$ level, just an order of magnitude beyond the current bound.

\section{Ho\v{r}ava-Lifshitz Gravity}

Ho\v{r}ava-Lifshitz (HL) gravity is an interesting suggestion for the ultraviolet (UV) completion of GR \citep{Horava:2009uw}, in which gravity turns out to be 
power-countable renormalizable at an UV fixed point. GR is supposed to be recovered at an infra-red (IR) fixed point, as the theory goes from high to low-energy scales. A renormalizable theory is obtained abandoning Lorentz symmetry at high-energies \citep{Horava:2009uw} through an anisotropic scaling between space and time: $\vec{r} \rightarrow b\vec{r}$ and $t \rightarrow b^{z} t$, $b$ being a scale parameter. The dynamical critical exponent $z$ is chosen so to ensure that the gravitational coupling constant is dimensionless, which makes possible a renormalizable interaction. As the Lorentz symmetry is recovered at the IR fixed point, $z$ flows to $z=1$ in this limit (see Refs. \citep{Kostelecky,Bluhn,bumblebee,BC2006} for other schemes to introduce Lorentz symmetry violation into the gravity sector). 

The anisotropy between space and time matches quite well within the $3+1$ Arnowitt-Deser-Misner (ADM) splitting formalism \citep{Arnowitt:1962hi} of the Hamiltonian formulation of GR. Following Ref. \citep{Horava:2009uw}, a foliation,  parametrized by a global time $t$, is introduced. Since the global diffeomorphism is not valid anymore, one imposes a weaker form of this symmetry, the so-called  {\it foliation-preserving diffeomorphism}. Choosing this approach, the lapse function, $N$, is constrained to be a function only of the time coordinate, $i.e.$ $N=N(t)$. This assumption satisfies the {\it projectability condition} \citep{Horava:2009uw}. The gravitational Lagrangian for this anisotropic scenario is obtained considering all relevant terms at an UV fixed point ($z\neq 1$) and, that at the IR fixed point, only $z=1$ terms survive and GR is presumably recovered. At this point, it is possible to consider the simplifying {\it detailed balance condition} \citep{Horava:2009uw,Horava2011} or not. However, given that the number of allowed terms in the action is not so large, the detailed balance Lagrangian can be obtained as particular case, actually through a proper choice of coefficients \citep{Sotiriou:2009gy}. 

Cosmological considerations have been extensively studied in the context of HL gravity (for a review, see Ref. \citep{Mukohyama:2010xz}) and we argue that quantum cosmology allows for acquiring some relevant insight about certain features of HL gravity 
\citep{Garattini:2009ns,BZ2011}. 

In order to obtain the minisuperspace Wheeler-DeWitt (WDW) equation the Robertson-Walker (RG) metric with
$R\times S^{3}$ topology is considered, 

\begin{equation}\label{eq:FLRWmetric}
 ds^{2}=\sigma^{2}\left(-N(t)^{2}dt^{2}+a^{2}\gamma_{ij}dx^{i}dx^{j}\right) ,
\end{equation}

\noindent where $i,j=1,2,3$, $\sigma^{2}$ is a normalization constant and $\gamma_{ij}$ is the metric of the unit $3$-sphere, 
$\gamma_{ij}=\mbox{diag}\left(\frac{1}{1-r^ {2}}, r^{2}, r^{2}\sin^{2}\theta \right)$.

The required extrinsic curvature takes the form:

\begin{equation}\label{eq:extrinsiccurvature}
 K_{ij}={1 \over 2\sigma N}\left(-{\partial g_{ij} \over \partial t}+\nabla_{i}N_{j}+\nabla_{j}N_{i}\right),
\end{equation}

\noindent where $N^{i}$ is the ADM shift vector and $\nabla_{i}$ denotes the $3$-dimensional covariant derivative. As $N_{i}=0$ for RW-like spaces,  
 
\begin{equation} \label{eq:Kij}
 K_{ij}=-{1 \over \sigma N} {\dot{a} \over a} g_{ij}.
\end{equation}

\noindent The Ricci components of the $3$-metric and its trace can also be obtained, as the foliation is a surface of maximum symmetry:

\begin{equation} \label{eq:3metric}
R_{ij}={2 \over \sigma^{2}a^{2}}g_{ij}~, \\
R={6 \over \sigma^{2}a^{2}}. 
\end{equation}

The action for the projectable HL gravity without detailed balance  is given by \citep{Sotiriou:2009gy}:
\begin{eqnarray}\label{eq:HLaction1}
&S_{HL}={M_{Pl}^{2} \over 2} \int d^{3}x dt N\sqrt{g} [K_{ij}K^{ij}-\lambda K^{2}   \nonumber \\
&-g_{0}M_{Pl}^{2}-g_{1}R -g_{2}{M_{Pl}}^{-2}R^{2}- \nonumber \\
&g_{3}{M_{Pl}}^{-2}R_{ij}R^{ij}-g_{4}{M_{Pl}}^{-4}R^{3}-g_{5}M_{Pl}^{-4}R(R^{i}_{\;j}R^{j}_{\;i}) \nonumber \\
&-g_{6}{M_{Pl}}^{-4}R^{i}_{\;j}R^{j}_{\;k}R^{k}_{\;i}  -g_{7}{M_{Pl}}^{-4}R\nabla^{2}R  \nonumber \\
&-g_{8}{M_{Pl}}^{-4}\nabla_{i}R_{jk}\nabla^{i}R^{jk}]~, \nonumber \\
\end{eqnarray} 
where $g_{i}$ are coupling constants. The time coordinate can be rescaled in order to set $g_{1}=-1$, the GR value. One also defines the cosmological constant $\Lambda$ as $2\Lambda=g_{0}M_{\mbox{\tiny Pl}}^{2}$. An important feature of the IR limit is the presence of the constant $\lambda$ on the kinetic part of the HL action. GR is recovered provided  $\lambda\rightarrow 1$ (corresponding to the full diffeomorphism invariance), however, for that $\lambda$ must be a running constant and the symmetry to lead to the $\lambda=1$ GR value is still to be identified. Phenomenological bounds suggest however, that the value of $\lambda$ is quite close to the GR value \citep{Sotiriou:2009bx}.

The HL minisuperspace Hamiltonian is obtained through a Legendre transformation and is given by:

\begin{equation}\label{eq:hamiltonian}
 H={1 \over 2}{N \over a}\left(-\Pi_{a}^{2}-g_{C}a^{2} +g_{\Lambda}a^{4}+g_{r}+\frac{g_{s}}{a^{2}}\right) ~,
\end{equation}
where the canonical conjugate momentum associated to $a$ is expressed as:

\begin{equation}\label{eq:canonicalmomentum}
 \Pi_{a}={\partial \mathcal{L} \over \partial \dot{a}}=-{a \over N}\dot{a} ~,
\end{equation}
and, following Ref. \citep{Maeda:2010ke}, the dimensionless coupling constants are redefined: 

\beqa\label{eq:dimensionlessconstants}
g_{C} &=& {2 \over 3\lambda-1}, \hspace{0.1 cm} g_{\Lambda} = {\Lambda M_{Pl}^{-2} \over 18\pi^{2}(3\lambda-1)^{2}},\\ \nonumber
g_{r} &=& 24\pi^{2} (3g_{2}+g_{3}), \\ \nonumber
g_{s} &=& 288\pi^{4}(3\lambda -1) (9g_{4}+3g_{5}+g_{6}).
\eeqa

In order to implement the quantum cosmology programme, the classical minisuperspace Hamiltonian is turned into an operator on which the so-called wave function of the universe is applied to \citep{DeWitt1967,Hartle1983}. Notice that the procedure of obtaining the minisuperspace can be regarded as a coset space dimensional reduction process \citep{BM1992}. 
This is a relevant point since there is no global diffeomorphism in HL gravity, just a foliation-preserving diffeomorphism \citep{Horava:2009uw}. This can be seen as the lapse function no longer depends on the space-time variables, as in GR, but now it depends only on the global time $N=N(t)$. This implies that the Hamiltonian constraint is not local, however this problem can be circumvented for an homogeneous metric, like Eq. (\ref{eq:FLRWmetric}), as the integration over space can be performed. The canonical quantization is obtained by promoting the canonical conjugate momentum into an operator, $i.e.$ $\Pi_{a}\mapsto -i\frac{d}{d a}$. Due to ambiguities in the operator ordering, one writes $\Pi_{a}^{2}=-\frac{1}{a^{p}}\frac{d}{d a}\left(a^{p}\frac{d}{d a}\right)$ \citep{Hartle1983}. 
The choice of $p$ does not modify the semiclassical analysis, hence $p=0$ is chosen and the WDW equation is written as

 \begin{equation}\label{eq:wdwequation}
  \left\{{d^{2} \over d a^{2}}-g_{C}a^{2} +g_{\Lambda}a^{4}+g_{r}+{g_{s} \over a^{2}}\right\}\Psi(a)=0.
 \end{equation}

This equation is similar to the one-dimensional Schr\"{o}dinger equation for $\hbar=1$ and a particle with $m=1/2$  with $E=0$ and potential 

\begin{equation}\label{eq:HLpotential}
V(a)=g_{C}a^{2}-g_{\Lambda}a^{4}-g_{r}-{g_{s} \over a^{2}}~,
\end{equation}
and from this identification many results can be obtained.

The salient features of this minisuperspace model are that the HL gravity terms are dominant on short distances, altering the behaviour of GR on these scales. The configurations for which the HL gravity new terms act as a ``potential barrier''  close to the singularity, $a=0$ can be suitably chosen \citep{BZ2011}.

The solutions of the WDW equation are obtained considering the deWitt boundary condition, which assumes that the wave function vanishes at the singularity. For $a\ll1$, corresponding to the very early universe when the HL gravity terms dominate, the wave function is an exponential, typical of a classically forbidden region. A quantum bound for the coefficient $g_{s}$ is found \citep{BZ2011}. 

For a vanishing cosmological constant, an exact solution can be obtained. In this case, the singularity is avoided due to quantum effects as the probability to reach the singularity  $a=0$ vanishes and $g_{r}$ is found to be quantized. Fixing the value of $g_{s}$, for large values of $n$ (large $g_{r}$ values), the wave function corresponds to a classical universe, in analogy to the correspondence principle of the old quantum mechanics \citep{BZ2011}. 

For the very late universe, the curvature and the cosmological constant terms dominate and, one finds that, for $\Lambda=0$ or $\Lambda<0$, the wave function is exponentially suppressed, denoting that this region is classically forbidden. For a positive cosmological constant case, it is found a damped oscillatory behaviour as in the usual quantum cosmology for GR. 

Finally, from the semiclassical analysis it is found a semiclassical solution oscillating nearby the classical solution. For $\Lambda>0$, this leads, at late times, to a de Sitter space-time, as expected from GR. 

We therefore conclude that the quantum cosmology of the HL gravity matches the expectations of a quantum gravity model for the very early universe, as it provides a hint for a solution of the singularity problem, at least for the $\Lambda=0$ case. In specific situations, the model also suggests that the GR behaviour is recovered at the semiclassical limit \citep{BZ2011}.

\section{Final remarks}

The underlying microscopic theory of gravity remains still an enigma. Recently, several fundamental issues have been raised about whether gravity is a fundamental interaction after all and whether more symmetries should be incorporated or some principles violated. In cosmology, a major progress would be achieved if the dichotomy between GR with dark components and alternative theories of gravity could be resolved. A violation of the equivalent principle at large scales would not allow for identifying unambiguously the theory of gravity, but it would signal the inadequacy of GR. Of course, the identification of states associated to dark matter and dark energy  in colliders (see e.g.  \citep{BR2008} and references therein) or other experiments (for information on the various experiments to detect dark matter see \citep{DM2011}) would be a major step forward. The same can be stated about unparticle states. In what concerns the issues raised by the idea that gravity is an emergent property or that gravity at high energies can be properly formulated without Lorentz symmetry, interesting concrete challenges are posed and these allow for  identifying the critical experimental tests that characterize these proposals. In any case, much is still to be learned about the inner process that connects the energy-momentum tensor and the space-time curvature as prescribed by GR, an issue that is getting more and more pressing and that indicates that there is still plenty of excitement ahead in the research of gravity.  
 
\begin{acknowledgements}
I  would like to thank Sonia Anton and Mariateresa Crosta for the superb organization of the GREAT meeting at Porto and for the kind invitation to deliver two lectures at a most interesting meeting.
\end{acknowledgements}

\bibliographystyle{aa}

\end{document}